\documentclass[conference,9pt]{IEEEtran}

\usepackage{subfigure, graphicx, units, multirow, cellspace, wrapfig, multicol,
textcomp,amsfonts,algorithmic,RobStd,amsmath,array}
\usepackage{booktabs}
\graphicspath{{images/}}

\def\thesubsectiondis{\thesectiondis\Alph{subsection}.}
\def\thesubsubsectiondis{\thesubsectiondis\arabic{subsubsection})}

\makeatletter
\def\section{\@startsection{section}{1}{\z@}{1.5ex plus 0.5ex minus 0.5ex}% V1.6 3.0ex from 3.5ex
{1ex plus 1ex minus 0ex}{\normalfont\normalsize\centering\scshape\bfseries}}%

\def\subsection{\@startsection{subsection}{2}{\z@}{1.5ex plus 0.5ex minus 0.5ex}%
{0.5ex plus .5ex minus 0ex}{\normalfont\normalsize\bfseries}}%

\def\subsubsection{\@startsection{subsubsection}{3}{\parindent}{0ex plus 0.1ex minus 0.1ex}%
{0ex}{\normalfont\normalsize\bfseries}}%
\def\paragraph{\@startsection{paragraph}{4}{2\parindent}{0ex plus 0.1ex minus 0.1ex}%
{0ex}{\normalfont\normalsize\bfseries}}%
\def\fnum@figure{Figure~\thefigure}
\makeatother

\let\oldbibliography\thebibliography
\renewcommand{\thebibliography}[1]{%
  \oldbibliography{#1}%
}

\renewcommand{\IEEEiedlistdecl}{%
    \setlength{\labelindent}{-1.0em}%
    \setlength{\itemindent}{1.5em}%
}

\HideNotes
\IEEEoverridecommandlockouts

\setblstr[0.5]{1.0}
\title{HERMES: A Hierarchical Broadcast-Based Silicon Photonic Interconnect for Scalable Many-Core Systems
}
\author{
\IEEEauthorblockN{\large  Moustafa~Mohamed\IEEEauthorrefmark{1}, Zheng~Li\IEEEauthorrefmark{2}, Xi~Chen\IEEEauthorrefmark{2} and Alan Mickelson\IEEEauthorrefmark{2}}
\IEEEauthorblockA{
\IEEEauthorrefmark{1} \small Department of Engineering Mathematics and Physics,
Cairo University, Giza, Egypt
}

\IEEEauthorblockA{
\IEEEauthorrefmark{2} \small Department of Electrical, Computer, and Energy Engineering,
University of Colorado, Boulder, CO~~80309, U.S.A
}
\IEEEauthorblockA{
\IEEEauthorrefmark{1}moustafa@eng.cu.edu.eg, \IEEEauthorrefmark{2}\{zheng.li, xi.chen, alan.mickelson\}@colorado.edu,
\vspace{-1\bls}
}
}
\normalsize

\begin{document}

%\abovedisplayskip=0.5\baselineskip
%\belowdisplayskip=0.5\baselineskip
\maketitle
%\pagestyle{empty}
%
%\BlankFootnote{ This paper was supported in part by the NSF under awards CCF-0829950,
%in part by the National Natural Science Foundation of China (NSFC) under grant
%\#60236020 and the Specialized Research Fund for the Doctoral Program of Higher
%Education (SRFDP) \#20050003083.
%
%~
%
%Permission to make digital or hard copies of all or part of this
%work for personal or classroom use is granted without fee provided that copies
%are not made or distributed for profit or commercial advantage and that copies
%bear this notice and the full citation on the first page.  To copy otherwise,
%to republish, to post on servers or to redistribute to lists, requires prior
%specific permission and/or a fee.\\
%DAC'09, July 26-31, 2009, San Francisco, California, USA.\\
%Copyright 2009 ACM 978-1-60558-497-3/09/07 ... \$5.00.
%}
%
%

\begin{abstract}
Optical interconnection networks, as enabled by recent advances in silicon
photonic device and fabrication technology, have the
potential to address on-chip and off-chip communication
bottlenecks in many-core systems. Although several designs have shown
superior power efficiency and performance compared to electrical
alternatives, these networks will not scale to the thousands of cores required
in the future.

In this paper, we introduce Hermes, a hybrid network composed of an
optimized broadcast for power-efficient low-latency global-scale
coordination and circuit-switch sub-networks for high-throughput data
delivery.  This network will scale for use in thousand core
chip systems. At the physical level, SoI-based adiabatic coupler
has been designed to provide low-loss and compact optical power
splitting. Based on the adiabatic coupler, a topology based on 2-ary
folded butterfly is designed to provide linear power division in a
thousand core layout with minimal cross-overs. To address the network
agility and provide for efficient use of optical bandwidth,
a flow control and routing mechanism is introduced to dynamically
allocate bandwidth and provide fairness usage of network resources. At
the system level, bloom filter-based filtering for localization of
communication are designed for reducing global traffic. In addition, a
novel greedy-based data and workload migration are leveraged to
increase the locality of communication in a NUCA (non-uniform cache
access) architecture.
First order analytic evaluation  results
%based on many-core benchmarks
have indicated that Hermes is scalable to at least 1024 cores and
offers significant performance improvement and power savings over
prior silicon photonic designs.

\end{abstract}

\section{Introduction}

%\Note{Zheng: Change the ``nanophotonic'' in title to ``silicon photonic''. Same
%thing to the rest of the paper.}

\Note{Sharad's five questions:}

\Note{What is the problem?}
With the advent of multi-core architectures, communication will become a
bottleneck for tomorrow's systems. Processing power is growing exponentially,
following Moore's law, through integration of more cores on-chip, putting more
pressure on global communication.  However, electrical interconnects have
failed to keep pace. Rather, according to International Technology Roadmap for
Semiconductors (ITRS) projections, electrical interconnects have evolved to be
power inefficient and a performance hindrance~\cite{itrs09}. Even looking at
future prospects of different system, we see high-bandwidth low-latency
networks as the gateway for high performance and power efficiency.
RAMCloud for
example proposes using DRAM as main storage for memory instead of hard-drives which would be used for backup. The
electrical interconnect path connecting processor to DRAM and processors to
each other will stand in the way of realization of RAMCloud for cloud computing~\cite{ousterhout09}.

The emerging technology of silicon photonics has been proposed as an
alternative solution to on-chip and off-chip communication~\cite{li10,
beamer10, koka10}.
%\Note{Replace light-speed with something related to WDM and
%high bandwidth.}
Not only does silicon photonics offer light-speed relay-free
communication and high bandwidth, in addition, it serves as an ultra-low-power network backplane.
As a demonstration of silicon photonics potential, several on-chip networks
have been proposed to solve on-chip communication problem~\cite{dana08, pan10,
joshi09may, sha08, cianchetti09jun}. Different topologies, flow control, and
routing mechanisms have been introduced. For instance, bus-based broadcast have
been shown to be superior to electrical alternatives~\cite{kirman06dec,
kurian10}. Crossbars were later introduced to further reduce power and improve
bandwidth of the system~\cite{dana08, pan10}. Hybrid electrical-optical
networks have been introduced to offer the arbitration mechanism and circumvent
bufferless routing in optical interconnects~\cite{cianchetti09jun, joshi09may,
sha08}. However, these architectures suffer from a common problem: scalability.
%Scalability is a crucial issue for large multi-core (and multi-chip) systems.
%For instance, power-scalability, a major challenge, prevents scaling these
%designs to thousand-core systems. Moreover, performance can not keep up with
%the ever-increasing bandwidth demand and latency constraints.
%Though they
%provide a bandwidth in the Terabit range, however, they are not scalable either
%performance-wise or power-wise.

\Note {Why aren't existing solutions good enough?}

%\Note{You didn't want to say ``we'' predict, rather it is  well-known}
We predict, along with several experts in the architecture domain, that future microprocessors will constitute tens to hundreds of
cores in the same package~\cite{koka10,borkar07,yeh08}.
%, with each composed of hundreds of
%cores~\cite{koka10,borkar07,yeh08}.
With such large scale systems current
silicon photonics designs are are not designed with such scale in mind and thus
do not provide the performance requirements nor satisfy the power constraints.
Bus-based broadcast topologies, for instance, exhibit an exponential power
growth with number of cores~\cite{kirman06dec,dana08}. Hybrid
optical-electrical networks suffer from the power inefficiency of electrical
interconnects~\cite{joshi09may,sha08,cianchetti09jun}.  Finally, the
widely-used serpentine-shaped crossbars exhibit a linearly increasing latency
and power with number of
cores~\cite{pan10,pan09jun,dana08}.
%\Note{it feels verbose of repeating these
%categories
%again in two consecutive paragraphs. Maybe remove the elaboration in the
%previous paragraph}

\Note {What is our solution?}
In this work, we propose Hermes, a hierarchical scalable silicon photonic
interconnect for large-scale systems, which addresses both performance and
power concerns for thousand core systems. The network is composed of:
\begin{itemize}
\item \textbf{Broadcast Network:} The network is a hybrid network consisting of
a novel high-bandwidth, low-latency optical broadcast network for arbitration
and communication. Unlike existing broadcast networks, this broadcast network
is scalable to large core-count with near ideal power division and delivery.
With such a broadcast structure one can improve the performance of the system
which is sensitive to latency of broadcast packets. Even if only a small
percentage of total traffic is broadcast, it can degrade the whole system
performance if not handled properly~\cite{jerger08jun}.

%\Note{Reviewer might not understand why you need a
%gloabl scale broadcast network. Need some motivations.}

\item \textbf{Linear Network:} The broadcast-network is augmented with a
circuit-switch optical network for long, throughput-hungry messages. This
linear point-to-point networK is designed to improve bandwidth for local
communication.

\item \textbf{Hierarchical communication domains:} The hierarchical approach
divides communication into local and global domains. This hierarchical approach
allows the scalability of the system to thousand core with minimal power
overhead. However, to enable efficient operation and high performance,
communication should be limited to local domains. Novel greedy-based data and
workload migration techniques to minimize communication in the global domain
and reduce latency. In addition, at the global-local interface we use a bloom
filter to filter requests that can be processed locally~\cite{udipi10}.

%The local and global domains have similar
%architecture, topology, flow control and routing mechanism. The main difference
%is in the number of waveguides (bandwidth) and length of communication path.
%In between the global and local domain, lies an electro-optic conversion point.

%\Note{Rather than explaining how the hirachy network is built, why not explain
%how hierachy helps scalability in terms of power and performance?}
% This
%interface is simply an optical-electrical conversion point and a bloom
%filter~\cite{udipi10}. This filter is customized to limit broadcast messages to
%filter out global communication thus reducing
%the overall traffic on the global domain.
%\item \textbf{System level optimizations:} Novel
%greedy-based data and workload migration techniques to minimize communication in
%the global domain and reduce latency. In addition, at the global-local interface
%we use a bloom filter to filter requests that can be processed
%locally~\cite{udipi10}. \Note{Isn't this part of the hierachy?}
\end{itemize}
%Local communication can be viewed as communication within the core and its
%dedicated cache, or between neighboring cores within same chip, or within the
%same chip which is the first level of hierarchy, or between neighboring
%chips.  Achieving locality at these different granularity can boost the
%performance of our network.

\Note{How good is our solution?}
Hermes shows superior performance and power efficiency compared to prior
silicon photonics designs~\cite{dana08, koka10,kurian10, li10}. At the
local communication domain, a linear-power scalability is achieved
as a result of the novel optimized broadcast network presented. Moreover, the
latency is minimized to match direct point to point latency.  As for bandwidth,
the circuit-switch linear network is optimized with that goal in mind. It
dynamically allocates bandwidth and allows high bandwidth point to point
communication for long cache-line messages. On the other hand,
at the global communication domain, latency is minimized by Hermes'
locality-improving techniques. First order evaluations show that Hermes can greatly improve the latency and bandwidth of the system
at low power level compared to other networks in the literature that are designed for tens of cores.
%Simulation has been conducted using real-world
%application traces such as PARSEC~\cite{bhadauria09} and SPECOMP~\cite{aslot01}
%running on a 1024 core system. The simulation study included Hermes and several
%optical networks proposed for thousand-core system and many-core systems. The
%simulation results demonstrate an improvement of ?? in latency and power and ??
%(should be finished when we have results).
%\Note{Need to mention we use PARSEC and realistic traces, also the 1024 core
%setup. Some of the sentence in the above paragraph could be move above to the
%item lists.}
%leveraged in this design.

%In overall, Hermes demonstrates a xx\% power saving over traditional bus-based
%networks~\cite{dana08, } and xx\% on average. Performance-wise, Hermes
%out-performed the other networks by xx\% in packet latency on average. On the other hand, to
%measure the effectiveness of the locality-improving techniques, we measured
%the suppressed global requests to show a xx\% locality improvement.

%\begin{comment}
This study  focuses on scalability of silicon photonic on-chip networks for
thousand-core systems, making the following contributions:
\begin{itemize}
\item	A novel hierarchical high-bandwidth and low-power network for communication in
future thousand-core systems.
%\item	A novel power-efficient, high-bandwidth, low-latency broadcast network
%which can scale to thousands of cores. The novelty lies in the new
%topology that leverages power splitters, such as adiabatic couplers, to evenly
%and with minimal loss distribute the power to different communicating nodes. Moreover,
%this network is general enough to fit other architectures that need broadcast or
%multi-cast for arbitration and/or communication.
%\item	Novel flow control and arbitration  to divide the bandwidth
%of the broadcast network among competing cores in a fair approach and with high
%utilization.  Moreover, it allows arbitration for the linear network with low
%latency.
\item	Silicon photonic is a new field and the design abstraction levels are
not well defined which lead to incorrect assumptions and designs. This study
addresses the design problem at different design levels including device-level,
physical-layout, topology, flow control, and system-level optimizations.
Moreover, we attempt to provide a device-level analysis for abstraction at
higher levels of design.
%\Note{Need to elaborate this point.}
%\item	A high-bandwidth circuit-switch linear network which is designed to improve
%performance for local communication. Moreover, its bandwidth is optimized for
%local point-to-point communication which typically requires high-bandwidth.
%\item	Novel local-communication-improving techniques through a low overhead
%high performance gains data and workload migration, and bloom filters.
%\item 	A simulation-based study showing that the proposed network at different
%levels of design can achieve high performance and power efficient communication
%that is required by multi-chip systems.
%\Note{I don't think simulation study is a
%contribuition.}
\end{itemize}
%\end{comment}
The rest of the paper is organized as follows:~\autoref{sec:background}
presents the background for this work.
%In ~\autoref{sec:motivation} we study
%the scalability of earlier silicon photonic network designs as a motivation for
%our approach to solve scalability in thousand-core systems.
Next, we discuss
the proposed design in~\autoref{sec:design}. In~\autoref{sec:results} we
describe our evaluation methodology and present our results
then we conclude
in~\autoref{sec:conclusion}.

\section{Background}
\label{sec:background}
Advances in silicon photonics have enabled integration of optical devices on a
chip with high density. The integration of silicon photonic devices has been
leveraged in on-chip interconnect to replace copper interconnects. In this
section we present a brief overview of the structure of silicon photonic
interconnects in general. In addition, we also discuss the design of thousand-core system
and how to overcome the main challenges facing its realization~\cite{koka10,borkar07,yeh08}.
%Finally, we
%present a survey of broadcast networks leveraged in several photonic silicon
%on-chip networks. \Note{I think you removed the survey part} We analyze them in terms of power, latency, and bandwidth.

\subsection{Basic components}
\label{sec:basicComp}
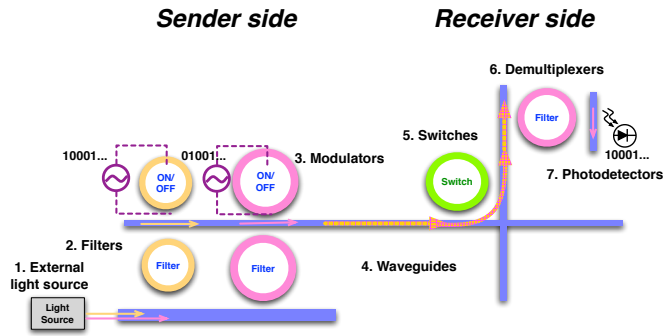
\ImageCap{optic}{1.0}{Optical path in silicon photonic networks\label{fig:opticalPath}}

The optical path in the silicon photonic network-on-chip mimics its counterpart
in telecommunication as shown in~\autoref{fig:opticalPath}. The optical path
starts with a laser-source which will be located off-chip into the foreseeable
future~\cite{bea08a}. The next stage is a filter that divides the broadband
optical light into different channels in the frequency domain~\cite{xiao07}.
These channels provide the necessary bandwidth through
wave-division-multiplexing (WDM) which waveguides cannot provide through
space-division-multiplexing (SDM). SDM is widely used by running multiple
waveguides in parallel to increase bandwidth. In the stage following filters,
the demultiplexed channels are modulated with the digital stream of data and
carried on a waveguide~\cite{reed10,gli10,guiliangli11,lee10}. The waveguide is
the conduit for the optical signal across the chip. Along the optical path, the
optical signal may encounter one or more switches that divert the path of the
signal~\cite{Xi10jul,liu09,biberman11,lee10}. In addition, it may encounter
electro-optic conversion points for signal regeneration. Finally, it reaches a
photodetector at the destination~\cite{zheng10jan,jurgen10,liao11}.

%\ImageCap{measurements}{1.0}{Measurement setup and fabricated
%devices\label{fig:setup}}

Individual silicon photonic components have been designed, fabricated, and
characterized.  The Emtnano team~\cite{emtnano} have fabricated in Epixfab~\cite{epixfab} several micro-ring
based passive filters, modulators and wide-band switches.
% as shown
%in~\autoref{fig:setup}.
These devices have been measured, characterized, and
modeled. The EMTNano team~\cite{emtnano} measurements demonstrate high quality factor and compact footprint
which makes it suitable for high density integration in optical on-chip
interconnect. Other components have been fabricated and characterized by other
groups such as thermo-optic micro-ring modulators having a power consumption
that ranges from \unit[7]{fJ/bit}~\cite{guiliangli11} while electro-optic
micro-rings have a power consumption of \unit[86]{fJ/bit}~\cite{chen09aug}
depending on the tuning techniques and switching speeds. Germanium-doped
silicon photo-detectors have been demonstrated exhibiting low power, as low as
\unit[33]{fJ/bit}~\cite{chen09aug}, and delivering over-GHz speed.

Based on the photonic devices, several groups have built optical links and
characterized them. For instance, Chen \etal demonstrated an optical link
operating at \unit[3]{Gbps} and dissipating as low as \unit[120]{fJ/bit} of
optical power.  Alduino \etal demonstrated a 4-channel WDM link operating at
\unit[10]{Gbps} with a maximum power dissipation of \unit[900]{fJ/bit} of
optical power~\cite{alduino10}. Finally, Zheng \etal built a low-power optical
link operating at \unit[10]{Gbps} and dissipating a few hundred fJ/bit of
optical power~\cite{zheng11}.

%However, at a larger scale
%mathematical models and simulators have been demonstrated~\cite{li11}. The
%models are based on characterization of fabricated devices upon which a
%library of T-matrices have been built which demonstrate high
%fidelity~\cite{chen10dec}.

\subsection{Thousand core system design}
%\Note{Discuss how to design a thousand core system. There are thermal and packaging challenges}
We envision that future systems will be many-core systems sharing the
same chip.
%Multiple conventional chips each about \unit[225]{mm$^2$},
The chip composed of hundreds to a thousand core and DRAM memory, will be integrated and
interconnected. Integration will be 3 dimensional to integrate the cores, DRAM
memory , and silicon photonics interconnect~\cite{black06}. At the local-level,
tens of chips are integrated and another layer of silicon photonics
interconnects run between the domains and connects them together. The whole system
can be a large distributed shared memory system enabled by our high bandwidth,
low latency photonic communication backplane. Memory will be shared and
distributed among processing and communication domains while caches located within each processing and communication domain will be shared
locally at the local-level.  Cache coherence for thousand core system is
maintained by snoopy cache coherence leveraging our novel broadcast network.
%enabled by our low latency, high bandwidth
%silicon photonic network.

Current silicon photonic networks do not scale to the thousand core architecture in latency, bandwidth, nor power. They are designed for low core count and do not account for problems arising in thousand core system as we show in~\autoref{sec:results}. This motivates our work presented here and our approach for on-chip network design in Hermes as we explain next.
%Multi-chip integration will be achieved through a large SOI (Silicon On
%Insulator) substrate that supports the different CMOS layers, DRAM layers, and
%the inter-chip silicon photonics layer. SOI substrate is used since silicon
%photonics requires an SOI substrate to confine light and minimize optical
%losses.

%Thermal management of this macrochip is a real challenge. This macrochip is
%expected to dissipate \unit[3.2]{kW} over an area of \unit[9$\times$9]{cm$^2$},
%such power levels requires liquid cooling as suggested by Koka \etal. Further
%details of the macrochip and thermal management are provided by Koka \etal's
%proposition~\cite{koka10}.

%\subsection{Thousand core system}
%One important aspect we consider in this discussion is CMOS-compatibility. Devices and designs that are not compatible with CMOS lithography are omitted.

%\input{motivation}

\section{Network design}
Current efforts in silicon photonics on-chip network design focus on
architecture techniques with limited investigation in the device level. The
abstraction of devices for silicon photonics on-chip network remains
incomplete. This introduces invalid assumptions or limited exploitation of
device capabilities. Herein, we attempt to bridge the gap between the device
level and architecture design by focusing on broadcast-based design and the
available power splitter options. More specifically, we give a device-level
review of different devices, their characteristics and their impact on the
architecture design. Then, we demonstrate how to exploit the capabilities of
the device at the architecture level.

In this section, we explain the construction of the broadcast network. We
follow a bottom-up approach starting from the power-splitter continuing upwards
to the network design.  We first compare between different alternatives for
power-splitting and justify our choice for the adiabatic coupler. Next, we show
the topology used for our broadcast network and how to place the different
components in the on-chip layout. Following that, we discuss the topology and
layout of the linear network. Finally, we explain how these networks interact
to provide the communication backplane for a thousand-core system.

\subsection{Power-Splitter}
\label{sec:powersplitter}
In a broadcast network, the power splitter plays a crucial role in splitting
the input power into equal beams. The device-level options vary in
characteristics. The main characteristics of concern are bandwidth,
insertion loss, CMOS compatibility, and splitting ratio of the output
beams. Next, we survey different options for power-splitting and focus on
the adiabatic coupler. %The adiabatic coupler  proves to be the best candidate
%for this job as we show in our discussion.  Simulation, fabrication, and
%measurement results support this conclusion.

\begin{table}
\caption{Different classes of silicon photonic networks\vspace{1\bls}}
\label{tab:device}
\centering
\begin{tabular}{l|cccc}
%\hline
\toprule
Device & Antenna~\cite{zhou10dec}	& Y \& Trench & Adiabatic\\
 &	& Splitter~\cite{tang10,qian08mar} & Coupler~\cite{cao10} \\
%&	& network & network & network\\
\hline
Bandwidth & 1.35-\unit[1.65]{$\mu$m}	& 1.5-\unit[1.6]{$\mu$m} & 1.5-\unit[1.6]{$\mu$m} \\
\hline
Power efficiency & 23\%	& 80\% & 96\% \\
\hline
Number of ports & 64 & Any & Any \\
\hline
CMOS Compatible	& Yes & No & Yes \\
\hline
Area & \unit[$5\times5$]{$\mu$m$^2$} & \unit[$11\times11$]{$\mu$m$^2$} & \unit[$200\times1.5$]{$\mu$m$^2$}\\
\hline
Process variation & Moderate & Sensitive & Insensitive\\
\bottomrule
\end{tabular}
%\vspace{-2\bls}
\end{table}

Power splitting can be achieved by several means as shown in~\autoref{tab:device}. For example, an antenna
array splits a single input to many outputs with a wide-wavelength-band
operation. However, it suffers from low bandwidth and power
inefficiency. A maximum of 64 ports can be designed and fabricated; beyond
that, power efficiency and bandwidth degrade to an unacceptable point as simulations indicate.
Moreover, the power efficiency of a 32-port antenna is 23\% for minimum
transmission. Consequently, the antenna array exhibits limited performance
and power efficiency which calls for further innovation in one-to-many power
splitters~\cite{zhou10dec}.

Y-splitters and trench-based splitters have been demonstrated as other options
for power splitting.  Y-splitters, on one hand, demonstrate sensitivity to fabrication
inaccuracies. The minute dimensions challenge photolithography
limits~\cite{tang10}. Hence, we exclude them from consideration in the near future. On
the other hand, trench-based splitters suffer from low power efficiency,
which can be as low as 80\%~\cite{qian07dec,qian08mar}.

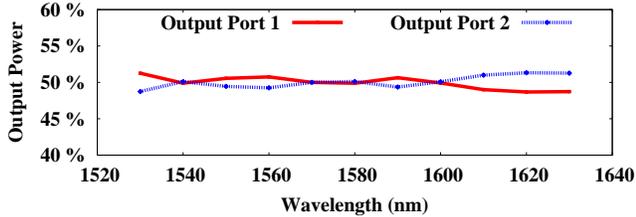
\ImageCap{acpower}{1.0}{Power splitting ratio of adiabatic coupler output ports\label{fig:acPower}}

Adiabatic coupler, the device chosen for building our network, combines several
advantages with low overhead:
\begin{itemize}
\item	\textbf{Bandwidth}: It has a wide wavelength band of operation that
extends beyond the C+L band (\unit[1530-1625]{nm}). This wide band of operation
adds another advantage which is immunity to thermal variations. Thermal variations that shift
the bandwidth, lead to insignificant impact on the
device characteristics due to its wide wavelength band.
\item	\textbf{Splitting Ratio:} The adiabatic coupler exhibits balanced
splitting ratio of the output beams. Simulation results indicate a 48-52\%
splitting ratio as shown in~\autoref{fig:acPower}. This result has been
confirmed by fabrication and measurements~\cite{cao10}.
%Our measurement results in our labs
%show ??.
\item	\textbf{Insertion loss}: The insertion loss of the adiabatic coupler is inherently negligible due
to the adiabatic nature of the device. The only loss worth mentioning is the
waveguide loss, which is \unit[2.5]{db/cm}~\cite{epixfab}
\item	\textbf{Number of ports:} The number of ports is 2 outputs but can be extended to any number of ports by cascading the device to form a network as we will show in Hermes broadcast network.
\item	\textbf{CMOS compatibility}: Cao \etal have successfully fabricated an adiabatic coupler in SOI and demonstrated the characteristics experimentally~\cite{cao10}. This confirms that adiabatic coupler can be successfully fabricated and integrated with CMOS process.
%designed and fabricated several designs of the adiabatic coupler in LETI in a
%\unit[200]{mm} SOI process~\cite{epixfab}.
\item	\textbf{Immunity to process variation} Our simulations
shows that for worst case process variation (10\% increase in waveguide width
accompanied by decrease in gap width) shows a 45-55\% power splitting across
the C+L band. Hence, the splitting ration and bandwidth remain at high levels despite the change in design dimensions.
\end{itemize}
%Measurement results show xx.

Another characteristic of adiabatic coupler that we have leveraged in our
network design is reciprocity of the device as our simulation results indicate,
which implies that the output and input ports can be exchanged while
maintaining the same power splitting functionality. Moreover, the adiabatic
coupler is a two-input, two-output device which makes it function as a
wavelength-combiner in addition to splitter. That means that if two input ports
have signals at different wavelengths, then the two signals will be split at
the two output ports as if they had been combined at the input.

%\ImageCap{ac}{1.0}{Microscope images of our fabricated adiabatic coupler\label{fig:ac}}

The only drawback of the adiabatic coupler compared to the other alternatives,
is its relatively long dimension, which may reach \unit[200]{$\mu$m}~\cite{cao10}
% as in our
%fabricated device in~\autoref{fig:ac}.
However, this drawback does not impact
our design, since up to sixteen adiabatic couplers are used on-chip where area
is scarce and this number can easily fit on-chip.
% and off-chip.

\subsection{Broadcast-network design}
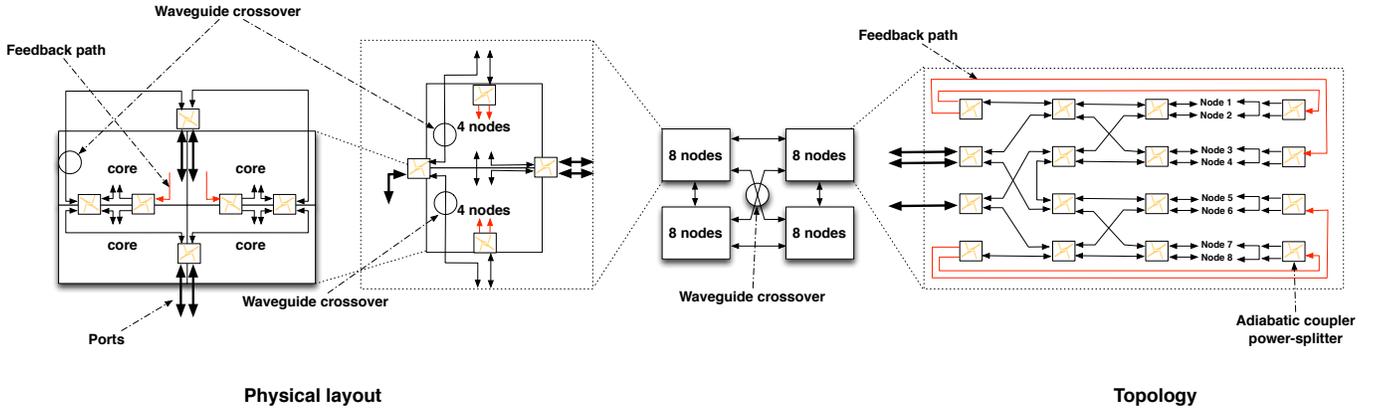
\ImageCapWide{hermes}{1.0}{Topology and physical layout of Hermes broadcast network\label{fig:hermes}}

In this section we demonstrate how to connect the adiabatic coupler in order to
achieve broadcast  from one node to the others in a performance- and power-
efficient manner. First, the topology is presented; following that, the layout is
explained. Finally, we extend the design from a single waveguide to a
high-bandwidth design, with an analysis of the involved trade-offs.

Our goal from the presented topology is to achieve a latency equivalent to
point-to-point connection while having a high bandwidth as close as to
bus and corssbar networks. Hermes broadcast can achieve a latency
equivalent to point to point latency, where waveguides connecting the far ends
of the chip run diagonally across layout. As for bandwidth, the bandwidth of
Hermes is higher than traditional networks and can achieve close to bus- and
crossbar- based bandwidth by running multiple waveguides in parallel. The only limitation
on number of parallel waveguides is crossovers. We show how to reduce crossover
loss in large bandwidth networks. These characteristics which we demonstrate in this section meets the ideal case we project in~\autoref{sec:performancescalability}.

The basic topology of the network is similar to a folded 2-ary butterfly
network with some adjustments as shown in~\autoref{fig:hermes}.  Each link in
the figure represents a two-way link. Each block in the diagram represents a
two-input two-output adiabatic coupler. Each coupler serves as both a
power-splitter (splits the input signal into two signal of equal length) and
wavelength-combiner (combines the two input signals from different inputs to same output). The
two-ary butterfly topology is folded network, carrying optical signal
in both directions. However, there are alternate feedback paths shown in red to
avoid the same wavelength traveling along the same optical link in different
directions which may lead to interference. The topology including the
butterfly topology and the alternate feedback paths form a power splitting
network of $\lceil \mathit{log}_2 \left(n\right) \rceil$ levels as the minimum
number of levels. It provides a near ideal-linear power division all-to-all
broadcast network.
%The
%extra levels included are due to the fact that an output port of an adiabatic
%coupler can not loop back as an input to the same port. Such a connection would
%lead to wavelength interference, since the same signal would be traveling back and
%forth on the same path.

The layout of the network shown in~\autoref{fig:hermes} depicts the placement
of the adiabatic coupler on-chip and the routing of waveguides between the
devices.  In this layout
%each neighboring nodes is connected via an adiabatic
%coupler.
each group of two processing cores connect to an adiabatic coupler
forming a cluster. Each two neighboring cluster connect to the next level of
the topology through another stage of adiabatic coupler forming a bigger
cluster.
%Mean feedback paths are provided to connect signals of nodes in
%same cluster.
After we reach a cluster of eight nodes these eight nodes are
connected to the rest of the clusters through direct connections encountering a single waveguide
crossover. In addition, feedback paths are provided as shown in red in~\autoref{fig:hermes} which also have a maximum of two waveguide crossover. Hence, the maximum number of waveguide crossover along any optical path in this network is one waveguide crossover.
%Each interconnected set of nodes is then hierarchically connected by another
%adiabatic coupler. After $\mathit{log}_2 \left(n\right) +1$ levels, the
%different groups, shown in~\autoref{fig:layout}, are connected through routing
%of the waveguides.

In order to achieve a high bandwidth network, more than one waveguide need to
run in parallel. However, such a network can introduce excess loss at
intersections. Hence, careful design of the intersections and number of
waveguides is necessary. In this design, we use the waveguide-crossings
designed by Popovic \etal~\cite{popovic07}. The loss in this structure has been
known to be as low as \unit[0.045]{dB/crossing}. While the crosstalk is
measured at \unit[-35]{dB}, this has a less pronounced impact on the system
compared to the loss; therefore, in our analysis we focus on the crossing loss.
For a hundred parallel waveguide, we get \unit[4.5]{dB} providing a plethora of
bandwidth with a small optical power penalty.

In case of high waveguide count, in order to reduce the number of crossovers in
the network one may opt for extending the number of adiabatic coupler levels by
more levels. This approach on one hand reduces the number of crossovers but
introduces a \unit[3]{dB} loss per extra level in the topology. Hence, a
tradeoff between crossover loss versus adiabatic coupler excess loss requires careful
design.

%The extra levels in the butterfly topology yields paths of higher
%power budget than others. This unbalance in power budget is an
%opportunity, to use paths of low power budget to links that have less waveguide
%crossings, usually close nodes. Other paths with higher power budgets can
%undergo more waveguide crossings, usually to further nodes. This is
%demonstrated in the layout shown in~\autoref{fig:layout}. Based on our layout,
%we have achieved a bandwidth of xxx waveguides which provides xxx wavelengths
%for communication. These numbers were computed based on a \unit[35]{dB} power
%budget, and \unit[-18.9]{dB} receiver sensitivity~\cite{zheng10jan}, below the
%nonlinear limit~\cite{TPA}.

\subsection{Linear network}
\label{sec:linear}

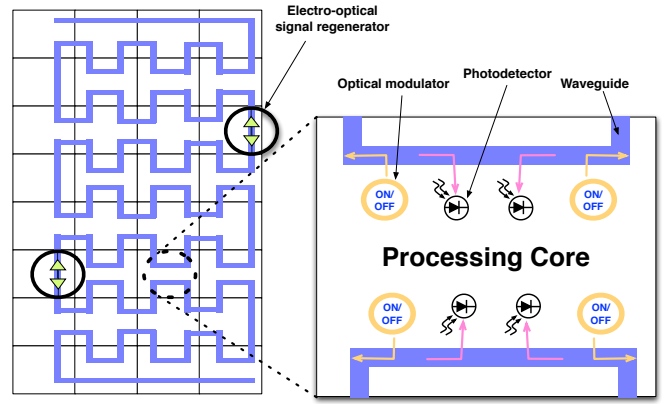
\ImageCap{linear}{1.0}{Linear circuit-switch network of Hermes\label{fig:linearLayout}}

In this section, we describe the second part of the network, the linear network.
This linear network serves as a medium for transferring long
messages between nodes. Unlike earlier serpentine networks,
this network in addition to providing point-to-point communication links, it favors local
communication over global communication.

The goal we had in mind while designing this network was having high-bandwidth
point-to-point communication links. Since, serpentine shaped networks have
highest bandwidth, we follow the same topology; however, we propose changes to
improve concurrency and overcome the high optical loss of these long
waveguides. First, we change the topology from simple serpentine to the one we
show in~\autoref{fig:linearLayout}. This topology gives more access points and
reduces the distance between further nodes and improve local communication
bandwidth. Hence, congestion is alleviated and concurrency is improved.
Moreover, we add electro-optic conversion points to mitigate the optical loss
and linearize it in terms of number of cores. Next, we discuss in detail the
physical design of the linear network.

The layout of the network is shown in~\autoref{fig:linearLayout}. The waveguide
runs in a serpentine shape across the chip crossing the same processing core
twice. This layout does not introduce any crossings and allows a large number
of parallel waveguides, improving the bandwidth of the network as required for
the large packets transferred on this medium.  Moreover, the network provides
four access points for communication per node as shown
in~\autoref{fig:linearLayout}. This increases concurrency because more than one
path exists between communicating nodes. Finally, the topology of this network
reduces the distance between neighboring nodes, favoring local communication
over long-distance ones. This kind of locality can be exploited using
kernel-level approaches leveraged in this design such as data
migration~\cite{kim03} and workload migration~\cite{lis11}.

%\Note{This is the first time you mention off-resonance coupling. Indeed, the
%first time you mention resonance... Any clue that the reviewer might understand
%that your implication?}
Traditional techniques of designing long waveguides suffer from excessive
waveguide losses and off-resonance coupling losses. Waveguide loss is
attributed to the silicon loss of the waveguide, while off-resonance coupling
loss occur when a non-resonant wavelength passes through the through port of
the filter, switch, or modulator.  To overcome these two limitations, we employ
electro-optic conversion points.  These points re-generate the signal by
converting it to an electrical then back to optical signals. However, only two
electro-optic points are employed introducing a small latency and power
overhead. In addition, the packets on this network exhibit less sensitivity to
latency and the optical power savings compensates the electrical power added.
With electro-optic conversion, the longest optical path is \unit[8]{cm} adding
\unit[20]{dB} of overall waveguide loss according to Epixfab fabrication
process~\cite{epixfab}. In addition, we leverage wide band switches that reduce
the number of micro-rings and off-resonance coupling the optical signal needs
to tolerate~\cite{Xi10jul}. In our system, the maximum off-resonance coupling
along any optical path is less than \unit[4]{dB}.

\subsection{Hierarchical design}
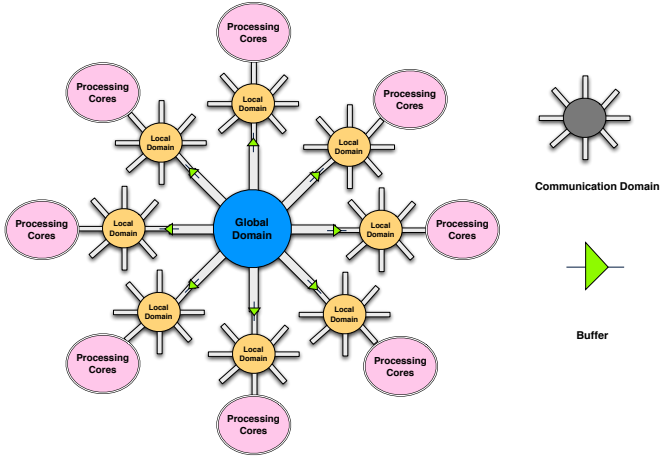
\ImageCap{hierarchy}{1.0}{Hierarchical Hermes silicon photonic network for a thousand core system\label{fig:hierarchy}}

In this section we extend the local-communication network (targeting 32 cores) to the global-communication scale
system that includes 1024 cores. In order to accommodate the performance and power requirements of the
system, we leverage a hierarchical approach, depicted in~\autoref{fig:hierarchy}. This hierarchical approach reduces the total power the system needs and improves performance. This improvement is two fold: First, the optical power demand and bandwidth requirement is limited to number of nodes in the small network rather than whole thousand core, second, by isolating communication in different sub-networks in the hierarchy we can improve power and performance.

The system is divided into local and global communication domains.  The
thousand-core system is divided into 32-core domains, and each domain contains 32 cores in order to have a balanced number of nodes at both the global and local domains. A
32-core is considered a single local communication domain. This
local domain is served by a single broadcast and linear network. The
thousand-core chip system represents another communication domain which is called a global
communication domain, where each local domain is considered to be a single node. The global
domain is served by a second hierarchy of network. This additional
hierarchical level is also composed of a broadcast and linear network. However,
the power and performance requirements of the two communication domains differ.
As shown in~\autoref{fig:hierarchy}, The the global domain (in blue) is connected through buffers to the local domains, while the local domains (in yellow) are connected to the cores (in pink).

The signal coming from one core to another core on another communication domain must pass
through the local and global communication domains. First, at the local domain,
there is a single access point to the global domain. At this access point the
electrical signal/data is converted to the optical domain. The signal must
travel from the source core to the global-local domain access-point. At the
global-local domain access point, the signal undergoes an electro-optic
conversion to regenerate the signal and buffer it, allowing it to travel the
long distance across the thousand-core system.  Next, the signal is carried in the
global communication domain to the destination communication domain. At the destination communication domain,
there is a single local-global domain access point to the core's local
communication domain. At this access point, the signal is re-generated and
transmitted to the destination core.

The main difference between the local and global domain networks lies in
bandwidth and traveling distance.  In the global communication domain, the
available area provides abundant space for space-division multiplexing. This
area allows a myriad of parallel-running waveguides which, in turn, increases
the bandwidth.
Moreover, the area in  the global domain
is more (a factor of 32X), which permits the increase in bandwidth of the broadcast network
since more waveguide-crossing can be accommodated. This increase in bandwidth
of the broadcast and linear network at the global communication domain allows
high concurrency and throughput which improves the performance of the whole
system despite the large number of communicating nodes. As for the long
traveling distance between communication domains, this leads to a quick increase in insertion
loss. Hence, to compensate for this loss, we add electro-optic conversion
points that regenerate the signal and allow it to travel a longer distance. In
our system we need a single regeneration point in the broadcast network and
eight regeneration points for the linear network in the global communication
domain.

The full network including broadcast and linear network at the local and global- communication level are
integrated with processing elements through 3D integration~\cite{black06}. The
network interface is connected to optical processing elements using TSV
(through silicon via).
%On the other hand, at the global-communication level, the global
%hierarchical domain of the network runs on different layers between the communication domains
%at both sides of the chip~\cite{koka10}.
%Hence, only nodes within the local domain know the
%local order and are not aware of any global order for the whole
%system~\cite{agarwal09}.

%The whole system is then connected globally by adding another level of hierarchy.
%Talk about the beam splitter - micro-rings - off-resonance coupling
%We are using TE mode only.
%The main components in the optical path.

\section{Flow control and system design}
\label{sec:design}
In this section we discuss the flow control of each sub-network of Hermes. We show how the
broadcast network is leveraged for multicast operations, thanks to its global
outreach. Meanwhile, the linear network serves as a point-to-point interconnect
for long messages.  These two sub-networks are connected hierarchically to serve
the large-scale system which comprises a thousand cores. Moreover, we show how
to improve locality at the system level to reduce global-communication and,
at the local-communication level, to improve linear network operation.

\subsection{Broadcast network flow control}
The broadcast network has global outreach which enables it to serve multicast
communication patterns. This kind of communication appears in cache coherence
protocol messages and in arbitration. Next we discuss how this all-to-all broadcast network can be used for arbitration and communication in the system.
%each communication role
%of the broadcast network in detail.

\subsubsection{Bandwidth division}
\label{sec:bw}
\ImageCap{BWDivision}{1.0}{Fair bandwidth division of the broadcast network\label{fig:bwFair}}

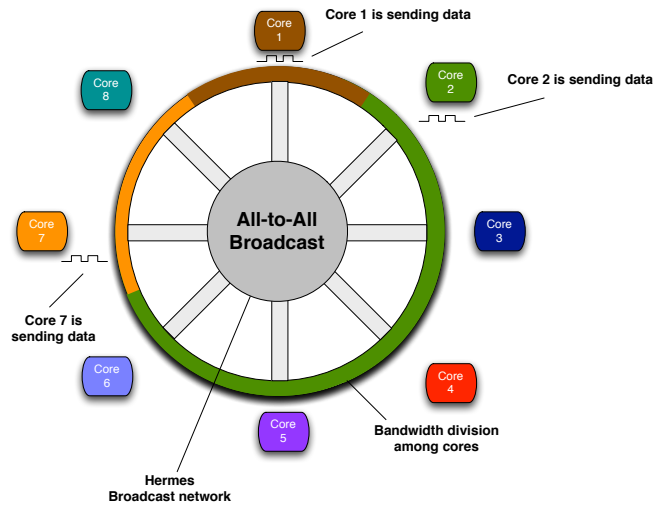
\ImageCap{BWDivisionUnfair}{1.0}{Efficient bandwidth division of the broadcast network\label{fig:bwEfficient}}
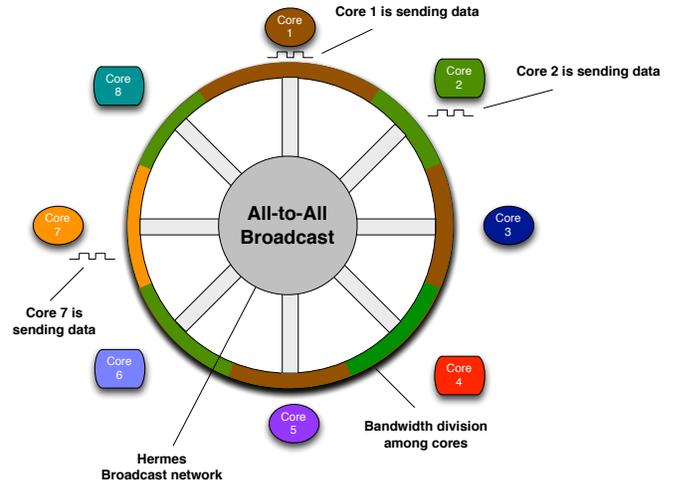
\ImageCap{BWDivisionAdaptive}{1.0}{Efficient and fair bandwidth division of the broadcast network\label{fig:bwAdaptive}}

The designed network offers a plethora of bandwidth. This bandwidth needs to be
shared efficiently and fairly between the communicating nodes. This section
describes how the bandwidth is divided between the communicating nodes. Unlike
earlier bandwidth division techniques, the presented approach herein is
superior from two perspectives: First it exploits the full bandwidth
dynamically and fairly even if contention is not 100\%. Second, it utilizes the
static laser power allocated to inactive nodes by sharing it among active
communicating nodes.

First, we consider a fair bandwidth division between nodes as shown in~\autoref{fig:bwFair}. In this figure, the bandwidth is equally divided between communicating nodes.
%Each node can only utilize its bandwidth when sending.
%For example in~\autoref{fig:bwFair}, core one, two, and seven are only sending data while the rest of the nodes are idle. Despite the fairness in bandwidth, the bandwidth utilization is less than 40\%.
To improve bandwidth utilization we propose another scheme, this can be achieved
by ordering the nodes in a circular closed set. Each node is assigned a dedicated
fair-share bandwidth that it can use for broadcast. In addition, it uses
the unused bandwidth of the next nodes in the ordered set until it encounters
a node that is sending. \autoref{fig:bwEfficient} shows the bandwidth division among
eight nodes. In this setting, nodes one, two, and seven  are sending on the network, while the
rest of the nodes are idle. In this case, the bandwidth between node two and
seven are free. Hence, node two can use this free bandwidth. Meanwhile, node seven has
one neighboring idle node whose bandwidth can be used. This scheme allows
efficient use of the overall bandwidth available to the network, which is 100\%.
Although bandwidth utilization is
optimized across the network, fairness in bandwidth sharing is not guaranteed.

The scheme above has a high utilization of bandwidth but unfortunately its
unfairness may lead to starvation. In order to improve the fairness of
bandwidth sharing, we divide the nodes into groups. Each group can share the
bandwidth as described above, but there is no cross-group sharing.
\autoref{fig:bwAdaptive} shows an example of fair bandwidth-sharing, in this setting
we have two groups which are group one including nodes one, three, five, and seven and  group two including nodes two, four, six, and eight. In Group One node one and seven
are sending while the rest are idle. The bandwidth of the nodes in this group
is shared between the group members as described earlier. In this case, node
one uses the bandwidth of nodes three and five. Meanwhile, node seven uses its bandwidth.
In Group Two, core two uses all the bandwidth in this group. This mechanism allows more fair sharing of bandwidth but
sacrifices some utilization efficiency of the bandwidth. This is because we may have groups with no nodes sending in it. Hence, there is a trade-off between fairness and utilization. To explain this,
consider $n$ nodes and $n$ groups. In that case each group (which is a node)
has a dedicated bandwidth. This achieves optimal fairness; however, since no
two groups may share bandwidth, the utilization of bandwidth drops.  On the other hand, when we have one group and $n$ nodes we have 100\% utilization and low fairness. By
adjusting the number of groups we can achieve a balance between fairness and
utilization efficiency.
%In this design we opt for ?? groups which through
%simulations has proven to give a good balance between utilization and fairness.

\subsubsection{Arbitration for the broadcast medium}
\label{sec:bArbitrate}
The broadcast network itself is a shared resource. In order to gain access to
the network, communicating nodes need to arbitrate for this resource with other
competing nodes and divide the bandwidth through a conflict resolution
mechanism.

\Note{Compare to other arbitration approaches. atomic coherence, token ring,
electrical assistance... Explain the arbitration design goal and challenges
briefly.}

A node can gain access to the broadcast medium by signaling the other nodes a
request for broadcast followed by sending data. For a $n$-node network, we
dedicate $n$-wavelengths of the bandwidth for signaling a request to start a
broadcast. At The next clock cycle, the requesting node may start broadcasting
directly. Thus, this flow control approach introduces a single cycle overhead
which is small. During the broadcast operation, the bandwidth is divided
between nodes as described in~\autoref{sec:bw}. By allowing several nodes to
broadcast at the same time, we improve concurrency, latency, and fairness of
the system.

Deadlock is not possible in the broadcast network since each node is given a
minimum amount of bandwidth no matter how many nodes are contending for
bandwidth. Moreover, the circuit-switching flow control inherit is silicon
photonics guarantees no contention once access to the network is satisfied.
Hence, deadlock will not occur in the broadcast network.
%Therefore the bandwidth
%needs to be divided between the sending nodes.

\subsubsection{Arbitration for the linear network}
\label{lArbitrate}
Arbitration for the linear network is performed through the broadcast network. The
procedure is a six phase process: (1) The requesting node checks a lookup
table to determine that the path is clear. (2) The requesting node signals the rest of the
nodes that a request for access has been made to the linear network. (3) The requesting node sends
the destination node address. (4) All the nodes update their
lookup table with the waveguide segments that are currently in use.
(6) When the requesting node completes transmission, it brings down the request
signal back to null indicating that it has completed the transaction. Thus, the
rest of the nodes may update the lookup table to clear the segments that were
reserved.

In order to guarantee fairness, a priority scheme is employed. Each node is
given a random priority and access to the linear network is given to node with
highest priority. This priority scheme shares the same seed; hence, no two
nodes will share same priority at any given time. Moreover, the priority scheme
changes periodically which avoids starvation situations. Moreover, this
guarantees that deadlock will not occur during arbitration since every node
will have a chance to get highest priority and gain access to the linear
network. Once the node gains access to the linear network, the path is reserved
and no cyclic wait over resources may occur. Hence, the network is deadlock
free during arbitration and communication~\cite{li10}.

In order to support such flow control, several resources are needed. First,
$n$-wavelengths are dedicated to signal a request to access the linear network.
Each of the $n$-nodes has a globally known wavelength. Another important
resource includes the lookup table. Each node is equipped with a lookup table to keep
track of the waveguide segments that are used and those that are free.  Thus, the lookup table frees
the broadcast network from contention for reserved resources. Finally, to
broadcast the destination address, each node consumes its dedicated bandwidth
as explained in~\autoref{sec:bw}.

%In the linear network neither deadlock nor starvation may occur during arbitration since the priority scheme changes and each node is given
\begin{comment}
\subsubsection{Coherence protocol support}
In cache coherence protocol part of the exchanged messages are multicast
messages. Without appropriate support of multicast, these types of messages
can severely degrade the performance of the network~\cite{jerger08jun}. The
broadcast network in this design solves this problem. It provides low-latency
medium for multicast messages.

In order for a node to send a multicast message, it arbitrates for the
broadcast network as explained in~\autoref{sec:bArbitrate}. Next, it computes
the available bandwidth after taking into account the shared bandwidth.
Following that, it transmits the message on the next clock cycle. Finally, it
terminates it request signal declaring the end of transmission and freeing its
bandwidth for other nodes to use.
\end{comment}

\subsubsection{Global order in broadcast network}

In snoopy protocol, global ordering becomes crucial to guarantee the
correctness of cache coherence protocol. Despite the fact that broadcast
provides global ordering naturally; however, the hierarchical design and
localization techniques (not all packets travel through the global domain)
complicates the situation. Even within the local communication domain, the
plethora of bandwidth allows multiple nodes to broadcast simultaneously. Hence,
global ordering requires extra effort. In this design, each node is given a
local order within its local domain. This order is random and changes
periodically which avoids starvation. Moreover, since all nodes share same
seed, collision will not occur. Packets sent are tagged by local order (5
bits). Packets received from different nodes are ordered by completion time,
then by node local order. In case of packets arriving from global domain, these
are sent by an intermediate node that receives packets from global domain and
sends it to local domain. In this case, these packets take the order of the node
that sends it. In the global domain, a similar scheme takes place where packets
received are ordered by an order local to the global domain
only~\cite{agarwal09}.
\begin{comment}
\subsection{Linear network flow control}
The linear network provides a high bandwidth and throughput network for the
long cache lines that needs to be transferred between communicating nodes. In
addition, it provides several paths for communication between the same nodes,
thus improving concurrency and throughput.

If a network needs to access the linear network it undergoes arbitration for
the path between the source and destination as describe
in~\autoref{fig:lArbitrate}. Following that, a point to point link is reserved
for communication between the source and destination. This path is leveraged
by the source node to send the cache line.

As discussed in~\autoref{sec:linear}, the topology of the linear network
favors local communication by reducing the distance between neighboring nodes.
In order to leverage this advantage we have employed communication
localization techniques. These techniques include workload
migration~\cite{lis11} and data migration~\cite{kim03}. These techniques have
improved the localization on-chip by xxx\% and have shown to be effective.
\end{comment}

\subsection{System-level optimizations}

At the system-level, the network design is hierarchical. Accessing
the global communication domain from the local communication domain
is done through an access point. All messages between these two
domains goes through the access point which has a dedicated minimal bandwidth to avoid performance bottlenecks. Re-direction of messages to the global domain depends on the message type. For
multicast messages, the access point will receive this message and using a
bloom filter it will determine whether this message should go to the global
domain or not, thus reducing the global domain traffic~\cite{broder04}. As for cache lines,
which are point to point communication, re-direction depends whether the destination is
on the same local-communication domain or not. If the destination is on another local-communication domain, the unicast
cache line is re-directed to the global domain. Otherwise no re-direction
occurs.

The cost of accessing the global communication domain is high in terms of power
and latency. Hence, at the system-level our goal is to keep communication
local. That is nodes that communicate often with each other are co-located in
the same local communication domain. To achieve this goal communication
localization techniques are leveraged. This includes workload
migration~\cite{lis11} and data migration~\cite{kim03}. In this section we
analytically formulate the thread-core assignment and page-core assignment in a
binary integer programming problem. Then, we propose a greedy algorithm that
gives near optimal results running at the kernel-level.

%These
%techniques have reduced the global traffic by xxx\%; thus greatly improving
%the performance of the overall system.

\subsubsection{Problem formulation}
\label{sec:problem}

The problem of assigning threads to cores to minimize communication in the
global communication domain can be formulated as an binary integer programming
problem. First, given $N$ threads and $M$ local-communication domain and a communication cost
$c_{ij}$ between thread $i$ and thread $j$ which is gathered from communication
statistics per thread upon each kernel switch, we want to find $x_{ijl}$ an
assignment of thread $i$ and thread $j$ to local-communication domain $l$ through workload migration.
Then we assign the shared memory pages to the local-communication domain using data migration.
$x_{ijl}$ is a binary variable which is one if thread $i$ and $j$ are in local-communication domain
$l$, while $c_{ij}$ is an integer weight. Our optimization problem can be
formulated as:
\begin{align}
Minimize: & \sum_{l=1}^{M} \sum_{i = 1}^{N} \sum_{j = i+1}^{n} c_{ij} x_{ijl}  & \label{equ:function} \\
subject \; to: & &\nonumber	\\
& A_{1} x = b_{1} & \label{equ:constraint1} \\
& A_{2} x = b_{2} &	\label{equ:constraint2} \\
& A_{3} x \le b_{3} &	\label{equ:constraint3}
\end{align}

where $x_{ijl}$ are the binary decision variables of size $\mathcal{O}\left(N^2
M\right)$ which determine which two threads are assigned to which local-communication domain.
~\autoref{equ:constraint1} is a set of constraints that ensure each local-communication domain has
exactly $N/M$ threads assigned to it where the number of constraints is
$\mathcal{O}\left(M\right)$, having one constraint per
local-communication domain.~\autoref{equ:constraint2} is a set of constraints that ensure each thread
is assigned with another $N/M - 1$ threads to the same local-communication domain, where the number
of constraints is $\mathcal{O}\left(N\right)$, having one constraint per
thread.~\autoref{equ:constraint3} ensures that each thread is assigned to one
local-communication domain only, where the number of constraints is $\mathcal{O}\left(N^3\right)$,
having one constraint per 3-thread-tuple (for each thread, no thread pair occur
in more than one local-communication domain).

The communication cost is computed as number of cache-lines exchanged between
each pair of threads. This can be easily counted using a counter per thread.
The kernel, periodically updates the communication cost for the whole system
upon every context switch after which the communication counter needs to be
reset.  Finally, after assigning threads to cores through workload migration,
the kernel assigns pages to local-communication domains through data migration~\cite{kim03}.

\subsubsection{Solution: Greedy algorithm}
Solving the binary integer programming problem presented
in~\autoref{sec:problem} is prohibitively expensive. It takes hours on a
desktop for small core counts (16 to 36 cores). Not to mention that binary
integer programming problems are NP-complete~\cite{garey90}. Hence, a more
efficient approach is needed. For this purpose in our system, we propose a
distributed greedy algorithm that re-assigns threads to cores. The algorithm is
presented in~\autoref{alg:greedy}. First the
$\mathit{GreedyMigration}\left(\right)$ function receive $\mathit{CommCost}$ as
the cost of communication between every pair of threads.  Then it heapifies the
matrix into a heap in a distributed manner~\cite{hija08}.  Next, we iteratively
extract the two clusters with highest communication and merge them in a greedy
manner.  Finally, we update the cost of communication of the different clusters
with the new cluster in parallel using multiple kernel threads.

\begin{algorithm}
\caption{$\mathit{GreedyMigration}\left(CommCost\right)$}
\label{alg:greedy}
{\small{
\begin{algorithmic}
\STATE \COMMENT{Create a heap of communication cost in parallel}
\STATE $\mathit{CommCostHeap}$ = $\mathit{ParallelHeapify}\left(\mathit{CommCost}\right)$
%\STATE Initialize(kernelThreads[]) \COMMENT{Initialize kernel threads}
%\FORALL[Iterate on all kernel-threads]{$kt_i$ in kernelThreads[]}
%\STATE	Initialize\left($kt_i$\right)
%\ENDFOR
\FORALL[For each  running thread]{$t_i \in \mathit{Threads[]}$}
\STATE $c_i$ = $\mathit{t}_i$ \COMMENT{Assign each thread to a cluster}
\ENDFOR
\WHILE[While heap not empty]{$\mathit{NOT}\;\mathit{CommCostHeap.Empty}\left(\right)$}
\STATE \COMMENT{Get the two clusters with highest communication}
\STATE $\left(i,j\right)$ = $\mathit{CommCostHeap.GetMax}\left(\right)$ %\COMMENT{Get two clusters with highest communication}
\STATE $c_k$ = $\mathit{Merge}\left(c_i,c_j\right)$ \COMMENT{Merge cluster $i$ and $j$ into $k$}
%\STATE $\mathit{Clusters.Remove}\left(c_i\right)$	\COMMENT{Remove cluster $i$}
%\STATE $\mathit{Clusters.Remove}\left(c_j\right)$	\COMMENT{Remove cluster $j$}
%\STATE $\mathit{Clusters.Add}\left(c_k\right)$	\COMMENT{Add new cluster $k$}
\FORALL[For each cluster]{$c_m$ $\in \mathit{Clusters}$ in parallel}
\STATE \COMMENT{Compute communication cost of $c_k$ with cluster $c_m$}
\STATE $\mathit{cost} =  \mathit{CommCostHeap.getCost} \left(c_m, c_i \right)$
\STATE $\mathit{cost} = cost+ \mathit{CommCostHeap.getCost} \left(c_m, c_j \right)$
\STATE	\COMMENT{Update CommCostHeap with new costs}
\STATE $\mathit{CommCostHeap.Add}\left(c_m, c_k, cost\right)$
%\STATE $\mathit{CommCostHeap.Remove}\left(c_m, c_i\right)$
%\STATE $\mathit{CommCostHeap.Remove}\left(c_m, c_j\right)$
\ENDFOR
\STATE $\mathit{Barrier()}$	\COMMENT{Synchronize all kernel threads}
\ENDWHILE
\end{algorithmic}
}}
\end{algorithm}

\section{Evaluation}
\label{sec:results}
%\Note{
%In this section we show how the power, and bandwidth of existing solution don't scale.
%Next we extrapolate and show that for a 1000 cores there is at least three level of hierarchies.

%We should talk about our solution and how it scales.
%Talk about our design top-down.
%}
In this section we evaluate the proposed design for a thousand-core system.
Using a first-order analytic evaluation, we compare the proposed design with state-
of-the-art solutions of silicon photonics scaled upto this a thousand-core system. First, we discuss the different networks we compare to in our simulations, and scale
them hierarchically to a thousand-core system in a similar fashion to Hermes.
Following this discussion, we present our performance results, then the power
results including a discussion of how Hermes was able to achieve superior
performance and power.
\begin{comment}
Studies on communication in multi-core architectures indicate that multicast
traffic if not handled efficiently, they can degrade the whole system
performance. These kind of packets are present in both snoopy and directory
cache coherence protocols. This indicates the importance of efficiently
supporting multicast operations with minimal overhead~\cite{jerger08jun}. We
suggest a broadcast medium for this job which has all-to-all outreach with
minimal latency.

In this section we motivate the importance of power-efficient, high bandwidth,
and low-latency broadcast networks proposed in Hermes. This in turn allows us
to have a scalable solution for future many-core systems. Herein, we discuss
the scalability limitations of various on-chip silicon photonic networks from the
performance and power aspects.
\end{comment}

We analyze the power, latency, and bandwidth scalability of multicast mechanism
for four classes of on-chip silicon photonic networks encompassing eight
state-of-the-art different networks as shown in~\autoref{tab:class}. The first
class is all-optical bus-based networks. This class includes networks that rely
solely on optics for communication. More specifically, the broadcast networks
in this class have a serpentine shape that reaches to all the cores (or core
cluster).  Examples of this class include Corona~\cite{dana08},
ATAC~\cite{kurian10} and Kirman's bus-shaped architecture~\cite{kir07}.  The
second class of networks is hybrid networks that mix optical and electrical
networks or routers such as Petracca \etal design~\cite{pet08} that leverages
an electrical network. Other networks such as Joshi \etal's Clos
network~\cite{joshi09may} and Phastlane~\cite{cianchetti09jun} leverage
electrical routers and optical links. The third class of networks have a
crossbar architecture where multiple nodes compete to send data to single
receiver. In this network, the waveguide takes a serpentine shape reaching out
to all nodes in the network. Examples of this class include
Corona~\cite{dana08} and Flexishare~\cite{pan10}. The fourth and final class
leverage antenna as a linear power division device for broadcast such as
Iris~\cite{li10}.

%\Note{An important point is missing here. Why we need broadcast, why do the
%reviewers need to concern about power/performance scalability of broadcast?}

%Nanophotonic interconnects, unlike electrical interconnects, has its unique
%characteristics. First, nanophotonic networks have ultra-low latency giving it
%an advanatage over the electrical alternative. Second, the power dissipation is
%within a few watts making it very power efficient and suitable for future
%power-limited systems. On the other hand, the challenging aspect of
%nanophotonics network design is its bufferless nature, two-dimensional planar
%layout, and its relatively expensive electrical to and from optical conversion
%power cost.

%There is a plethora of work on on-chip nanophotonic network design that attempt
%to leverage the advantages of nanophotonics, however, they suffer from similar
%problems that limit its scalability. For small systems, less than xxx
%cores, these systems give superb performance but beyond that performance and
%power efficiency deteriorate significantly making them unsuitable.

\subsection{Power scalability}
\label{sec:powerscalability}

\begin{table}
\caption{Different classes of silicon photonic networks\vspace{1\bls}}
\label{tab:class}
\centering
\begin{tabular}{|l|c|c|c|c|}
\hline
Network & Bus	& Hybrid & Crossbar & Antenna-based \\
%&	& network & network & network\\
\hline
Kirman Bus~\cite{kir07} &	$\surd$ & & & \\
\hline
ATAC~\cite{kurian10} &	$\surd$ & & & \\
\hline
Flexishare~\cite{pan10} & & & $\surd$ & \\
\hline
Corona~\cite{dana08} &	$\surd$ & & $\surd$ & \\
\hline
Phastlane~\cite{cianchetti09jun} & &  $\surd$  & & \\
\hline
Clos~\cite{joshi09may} & & $\surd$  & & \\
\hline
Columbia Mesh~\cite{pet08} & & $\surd$ & & \\
\hline
Iris~\cite{li10} & & &  & $\surd$ \\
\hline
\end{tabular}
%\vspace{-2\bls}
\end{table}

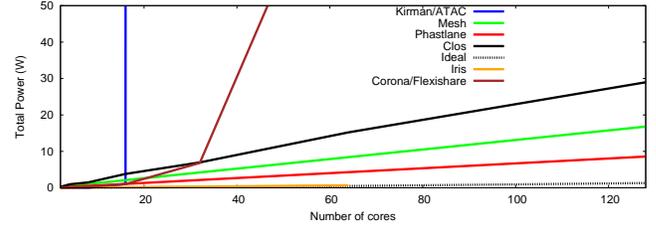
\ImageCap{powerscalability}{1.0}{Power scalability of silicon photonic networks\label{fig:powerscalability}}

%\Note{Can you labelled the figure with the four classes, it's kind of dizzy for
%people to remember all the network names, classifications and match it with the
%figure.}

In this section we analyze the power scalability of the four classes of on-chip
silicon photonic networks. In the first class of all-optical networks,  that have serpentine-shaped
topologies, these rely heavily on a broadcast network for arbitration and in
some cases for cache coherence packets. %\Note{I don't think Atomic Coherence has
%to use a broadcast network.}
The broadcast network is composed of a
bus topology where each communicating node taps half the available power
through a beam splitter. In this structure, half of the power goes to the core,
and the other half continues onto the rest of the cores. Linear power division
in bus-based broadcast, even though theoretically might be feasible, has not been demonstrated in the literature
%Power splitters that tap
%pre-determined power ratios have not been demonstrated in the literature (except \unit[3]{dB} couplers)
and their sensitivity to process variations remain unknown to our knowledge.
%\Note{why does broadband matter here to the
%reviewer?}
Reliable and process-variation immune power
splitters that are available in silicon photonics are \unit[3]{dB} power
splitters (50\% splitting ratios) as we have shown in~\autoref{sec:powersplitter}.
% Later on, we will provide a survey of power
%splitting devices and provide a more detailed comparison.
Moreover, to our
knowledge, prior work assuming a photonic bus do not provide details about the
device-level implementation of the bus-based architecture. The drawback of this
topology is the exponential growth of optical power with numbers of cores
($\mathcal{O} \left(2^N\right)$, where $N$ is number of cores). At a large
number of cores, the power consumption of the broadcast network dominates and
degrades the power efficiency of the overall system. As shown
in~\autoref{fig:powerscalability}, the power levels of the network is
acceptable for small networks. However, as the number of communicating nodes
exceed sixteen, the power consumption increases dramatically as one can see in
Bus in~\autoref{fig:powerscalability}. This makes the system unreliable since
the majority of the power is converted to heat~\cite{abdollahi09}.

In the second class of networks, the hybrid optical-electrical solutions suffer
from the high power consumption of the electrical components as one can see in~\autoref{fig:powerscalability}. Clos~\cite{joshi09may} has the highest power levels due to the high radix of its routers which greatly increases the power. Mesh by Columbia~\cite{pet08} has the second highest level because of the electrical links in addition to the routers in the electrical network. The overhead of electrical links have exceeded the power of buffered networks in Phastlane~\cite{cianchetti09jun} despite its bufferless routers. Finally, Phastlane~\cite{cianchetti09jun} has the lowest power due to the low radix routers and optical links.
Despite the linear scaling trend of power ($\mathcal{O} \left(N \right)$, where $N$ is number of cores)
in this class of networks, the total power consumption is relatively
high.
As shown in~\autoref{fig:powerscalability}, the power levels of this
class of networks is large --- even for small networks. For a 32 node network, the electrical power can reach ten watts.

The third class of networks have a crossbar topology. In this class, the
crossbar spans the whole cores in a serpentine-shaped topology. This serpentine
shape has a high optical loss which increases exponentially with number of
node, this can be explained as follow. Each node can send at any given time,
hence, each node has its own power source. Moreover, the waveguide length
increases with number of nodes but the length is longer than typical networks
and can reach tens of centimeters. Assuming a \unit[2.5]{dB/cm}~\cite{epixfab}
waveguide loss as we do assume for all other networks, then the total loss will
increase exponentially with number of nodes ($\mathcal{O} \left(10^{\sqrt{N}}
\right)$, where $N$ is number of cores).

The fourth and final class leverage an optical antenna for linear power division~\cite{zhou10dec}. The
main advantage of this approach is linear power division as shown
in~\autoref{fig:powerscalability}. However, there are two drawbacks: First,
optical antenna have poor power efficiency. The efficiency is as low as
23\%~\cite{zhou10dec} which limits the power efficiency of the statically
allocated optical power in the whole system. Moreover, the number of ports in
the antenna do not scale beyond 64 ports. Hence, the maximum system size is a
64-core system~\cite{li10}. %\Note{Does antenna also has cross issues?}

An ideal case is to have an optical network with linear power trends in the
number of cores ($\mathcal{O}\left(N\right)$, where $N$ is number of cores)
like Iris~\cite{li10} but with high optical efficiency. As we will show later,
Hermes can achieve this goal. Hermes provides linear power division at an
efficiency of 96\%.  This makes it superior to existing solutions and
attractive for managing multicast communication in on-chip networks. Moreover, at
large core count, and through hierarchical design the power is reduced to $\mathcal{O} \left(\sqrt{N}\right)$.

\subsection{Performance scalability}
\label{sec:performancescalability}
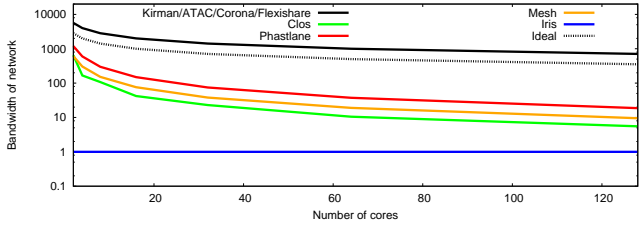
\ImageCap{bwscalability}{1.0}{Bandwidth scalability of silicon photonic networks\label{fig:bwscalability}}
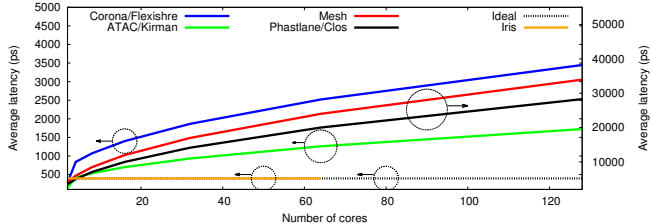
\ImageCap{latencyscalability}{1.0}{Latency scalability of silicon photonic networks\label{fig:latencyscalability}}

In this section we study the performance scalability of various networks.  More
specifically, we study how the bandwidth and latency of the broadcast network
scales with the number of cores per chip.  We define bandwidth as the bandwidth
available for a core under zero load condition and \unit[5]{W} power budget for
the electrical components.  On the other hand, we define latency as the worst
case traversal time for a packet under zero load.

%\Note{Will putting a table here much better for comparision?}

~\autoref{fig:bwscalability} demonstrates the scalability of bandwidth for
different designs. The first and third class represented in
Kirman/ATAC/Corona/Flexishare in~\autoref{fig:bwscalability} has a sub-linear
decay in bandwidth per core ($\mathcal{O} \left( \frac{1}{\sqrt{N}}\right)$,
where $N$ is number of cores).  This returns to the constant chip area
projected by ITRS~\cite{itrs09} which limits the bandwidth per core. However,
the bandwidth level are highest since the whole area is utilized for
waveguides. % it becomes comparable to other alternatives. As for
%the second class, the limited number of ports per antenna limits the total
%bandwidth. Hence, these designs have a very low bandwidth even for
The bandwidth in the second class is limited by the electrical power. As the
bandwidth increases, the link and/or router power increase. Under the power
constraint, the number of electrical components are greatly reduced and the
trend is inversely proportional to the number of cores ($\mathcal{O}
\left(1/N\right)$, where $N$ is number of cores) as shown in
the~\autoref{fig:bwscalability}.  We can see that
Phastlane~\cite{cianchetti09jun} has more bandwidth than Mesh~\cite{pet08}
which has more bandwidth than Clos~\cite{joshi09may}. This is because the
available bandwidth scales inversely with the power consumption. Under a fixed
power constraint low-power consuming networks can offer higher bandwidth
levels.  The fourth class of networks which is represented by Iris~\cite{li10}
has a constant bandwidth ($\mathcal{O}\left( 1 \right)$) and it is small due to
the limitation on number of ports in the antenna. The bandwidth available in
any clock cycle is 64 channels.  Finally, the ideal case would follow the same
trend of high bandwidth like class one and three (Bus and Crossbar). Hermes,
shows high bandwidth that follows the same trend as class one and three but at
a slightly lower bandwidth level.

On the other hand, ~\autoref{fig:latencyscalability} shows the scaling of
latency of the broadcast network in different designs. Class one and three of networks
have a high latency, linear in core-count ($\mathcal{O} \left(N\right)$, where
$N$ is number of cores), due to the serpentine shape of the broadcast network
which has to pass through all nodes in the network serially.  We can see that
Corona~\cite{dana08} and Flexishare~\cite{pan10} exhibit higher latency than
ATAC~\cite{kurian10} and Kirman~\cite{kir07} since there is an arbitration
cycle involved before sending the data. Class four, Iris~\cite{li10},
has constant and low latency, independent of the number of cores ($\mathcal{O}
\left(1\right)$), because the waveguide runs diagonally through the chip. This
is the optimal case having lowest latency. However, Iris~\cite{li10} does not
scale beyond 64 cores due to the limitation of number of ports in the antenna.
Class two (Mesh~\cite{pet08}, Phastlane~\cite{cianchetti09jun}, and
Clos~\cite{joshi09may}) has a sub-linear latency trend ($\mathcal{O} \left(
\sqrt{N} \right)$, where $N$ is number of cores), but exhibits high latency
levels due to the routing and switching overhead of the electrical routers.
Mesh~\cite{pet08} has a lower latency because it leverages bufferless routers,
meanwhile, Phastlane~\cite{cianchetti09jun} and Clos~\cite{joshi09may} use
buffered routers with higher delay. Finally, the ideal case should have low latency like Iris~\cite{li10} but scalable to large core count.
Hermes can offer this low latency and scale it to large core count. Moreover, through
the use of hierarchical design, the latency is reduced to $\mathcal{O} \left(\sqrt{N}\right)$

\section{Conclusion}
\label{sec:conclusion}
In this study we have presented a novel broadcast-based network that can
achieve linear power scalability with the number of nodes. Moreover, we have
scaled the network to a hierarchical network that can serve a thousand-core
chip system. First order power and performance evaluation of the proposed network
show superior results compared to state-of-the-art silicon photonics networks.
Moreover, communication locality have been greatly improved through
kernel-level workload and data migration.

%\setblstr{0.84}
%\section*{Acknowledgment}
%The authors would like to thank
%Dr. Pieter Dumon from Ghent University - IMEC,
%for providing valuable information about process variations of IMEC's process.
%Dr. Zheng Li for his comments on the paper.

\setblstr{0.9}
{\small
\bibliographystyle{IEEEtran}
\bibliography{robbib,nanodesign}
}
\end{document}